\documentclass[prl,aps,superscriptaddress,twocolumn,notitlepage,showpacs]{revtex4-1}
\usepackage{latexsym}
\usepackage{amsmath, dsfont}
\usepackage{amssymb}
\usepackage{graphicx}
\usepackage{caption}
\usepackage{physics}
\usepackage{subfigure}
\usepackage{float}
\usepackage{calrsfs}
\usepackage{mathrsfs}
\usepackage{notes2bib}
\usepackage{color}
\usepackage{txfonts}
\usepackage[justification=centering,
            format=plain]{caption}

\renewcommand{\raggedright}{\leftskip=0pt \rightskip=0pt plus 0cm}

\begin{document}

\title{Quantum amplification of spin currents in cavity magnonics by a parametric drive induced long-lived mode}

\author{Debsuvra Mukhopadhyay}
\email{debsosu16@tamu.edu}
\affiliation{Institute for Quantum Science and Engineering, Department of Physics and Astronomy,Texas A$\&$M University, College Station, TX 77843, USA}

\author{Jayakrishnan M. P. Nair}
\email{jayakrishnan00213@tamu.edu}
\affiliation{Institute for Quantum Science and Engineering, Department of Physics and Astronomy,Texas A$\&$M University, College Station, TX 77843, USA}

\author{G. S. Agarwal}
\email{girish.agarwal@tamu.edu}
\affiliation{Institute for Quantum Science and Engineering, Department of Physics and Astronomy,Texas A$\&$M University, College Station, TX 77843, USA}
\affiliation{Department of Biological and Agricultural Engineering, Texas A$\&$M University, College Station, TX 77843, USA}

\date{\today}

\begin{abstract}
Cavity-mediated magnon-magnon coupling can lead to a transfer of spin-wave excitations between two spatially separated magnetic samples. We enunciate how the application of a two-photon parametric drive to the cavity can lead to stark amplification in this transfer efficiency. The recurrent multiphoton absorption by the cavity opens up an infinite ladder of accessible energy levels, which can induce higher-order transitions within the magnon Fock space. This is reflected in a heightened spin-current response from one of the magnetic samples when the neighboring sample is coherently pumped. The enhancement induced by the parametric drive can be considerably high within the stable dynamical region. Specifically, near the periphery of the stability boundary, the spin current is amplified by several orders of magnitude. Such striking enhancement factors are attributed to the emergence of parametrically induced strong coherences precipitated by a long-lived mode. While contextualized in magnonics, the generality of the principle would allow applications to energy transfer between systems contained in parametric cavities.
\end{abstract}

\maketitle




Photon mediated interactions are a quintessential resource in various branches of sciences. A prime example is the well known dipole-dipole interaction (DDI) , which, according to quantum electrodynamics, arises from the exchange of a photon \cite{agarwal2012quantum} between two atoms. The DDI consists, in general, of a dissipative component as well which can vanish at certain separation of the dipoles. This DDI determines the energy/excitation transfer \cite{forster1948intermolecular, jones2019resonance} between, say, the donor and the acceptor molecules, and is of paramount importance in the generation of quantum gates \cite{nielsen2002quantum, divincenzo1995quantum}. It is therefore desirable to have a mechanism which can control and potentially improve such an interaction. One routine technique is to use high-quality cavities and employ large coupling between the cavity photons \cite{schafer2019modification, coles2014polariton, yuen2019polariton}. A natural question then arises - can one further improve the cavity-mediated energy transfer which would instrumental to the realization of quantum-enhanced fundamental interactions and development of sophisticated quantum machines and networks?

In this letter, we provide a definite answer to this question and demonstrate the possibility of enhancing photon-mediated transfer of excitations. Guided by the developments in quantum metrology using squeezed states \cite{PhysRevD.23.1693, unruh1983quantum, lawrie2019quantum, PhysRevResearch.4.L012014,taylor2016quantum}, we propose to use parametric interactions, which give rise to squeezed states of matter and light, to enhance the cavity-mediated DDI. We would specifically apply the idea in the context of cavity magnonics \cite{PhysRevLett.111.127003, PhysRevLett.113.083603, PhysRevLett.113.156401, PhysRevLett.120.057202, lachance2019hybrid, kostylev2016superstrong} and show significant amplification of the photon-mediated transfer of spin currents between magnetic samples \cite{PhysRevLett.118.217201} by using parametric interactions in cavities. We note that many other applications of parametric interactions have appeared in literature: enhanced cooling \cite{huang2009enhancement}; exponentially enhanced spin-cavity photon coupling \cite{PhysRevLett.120.093602, PhysRevLett.120.093601}; enhanced phonon-mediated spin-spin coupling in a system of spins coupled to a cantilever \cite{PhysRevLett.125.153602}; possibility of first order superradiant phase transitions \cite{zhu2020squeezed}; enhancement in the generation of entangling gates \cite{burd2021quantum}; amplification of small displacements of trapped ions \cite{burd2019quantum}.   

Here, we focus on cavity magnonics involving the coupling of high-quality microcavities and YIG spheres. These systems are attracting increasing attention \cite{PhysRevLett.114.227201, zhang2015magnon, chumak2015magnon} as favorable candidates to observe various semiclassical \cite{PhysRevB.93.174427, PhysRevB.105.064405, zhang2017observation, PhysRevLett.121.137203, PhysRevB.99.134426, PhysRevB.100.094415, PhysRevLett.123.127202, PhysRevLett.123.227201, PhysRevB.103.224401} and quantum phenomena \cite{tabuchi2015coherent, zhang2016cavity, PhysRevLett.121.203601, li2019entangling} at the macroscopic level. Some of the key developments include the coupling of magnons to a superconducting qubit \cite{tabuchi2015coherent} and phonons \cite{zhang2016cavity}, microwave-to-optical interconversion \cite{PhysRevB.93.174427, PhysRevB.105.064405}, exceptional points \cite{zhang2017observation}, entanglement \cite{PhysRevLett.121.203601, li2019entangling} and many more. One of the remarkable signatures of such a coupling was the observation of a cavity-mediated transfer of spin excitations \cite{PhysRevLett.118.217201}. Two YIG samples were placed at the opposite ends of a microwave cavity and by manipulating the cooperativity of one of them, researchers could detect the modifications in magnon population, namely, the spin current of the other. Here we demonstrate how the photon-mediated transfer of spin-wave excitations can be significantly boosted by a two-photon parametric drive applied to the cavity. The parametric interaction could be produced either from a $\chi^{(2)}$-type or $\chi^{(3)}$-type nonlinearity. By coherently pumping one of the samples, we probe modifications to the steady-state magnon occupancy in the neighboring sample as a function of the parametric drive strength. Our analysis showcases the emergence of parametrically induced coherences characterized by a long-lived mode leading to precipitous enhancement in the spin current response. The enhancement can be extraordinarily large around the two-photon resonance condition $\omega_d=\omega_p/2$, where $\omega_d$ and $\omega_p$ are the frequencies of the magnon drive and the parametric pump field respectively. This phenomenon is brought to bear by the cumulative effect of two-photon-excitation events. The cascaded absorption of photons by the concerned magnon mode opens up higher-order transition pathways within the magnon Fock space. The fact that magnificent enhancement factors can be achieved even within the permissible stable regime of the nonlinear dynamics underscores the utility of this scheme. The analysis presented in this letter is generic, pertinent to a wide class of systems with a parametrically driven component. This is because the DDI is ubiquitous. The investigated paradigm would apply, for example, to nonlinear Kerr boson systems driven far from equilibrium \cite{PhysRevB.105.245310}.

To set the stage, we shortly recapitulate the problem of  two spatially separated macroscopic ferrite samples of YIG coupled to the microcavity field. Owing to an effective cavity-mediated coupling between the two spheres, an external driving field applied to the first YIG sample would elicit a spin current from the second one. The system Hamiltonian, in the reference frame of the driving field, assumes the form \cite{PhysRevLett.120.057202, PhysRevB.94.224410}
\begin{align}
\mathcal{H}_0/\hbar=\Delta_ca^{\dagger}a+\sum_{j=1}^2\Delta_jm_j^{\dagger}m_j+\sum_{j=1}^2g_j[a^{\dagger}m_j+am^{\dagger}]\notag\\+i\Omega(m_1^{\dagger}-m_1),
\end{align}
where $a$ is the annihilation operator representing the cavity, $\Delta_c=\omega_c-\omega_d$ is the cavity detuning, $m_1,m_2$ are the two Kittel modes representing magnonic excitations in the two samples, and $\Delta_1=\omega_1-\omega_d$, $\Delta_2=\omega_2-\omega_d$ the respective detunings. For each $j$, the parameter $g_j=(\sqrt{5}/2)\gamma_e\sqrt{N_j}B_{\text{vac}}$ denotes the coherent magnon-photon interaction strength, with $\gamma_e$ being the gyromagnetic ratio, $B_{\text{vac}}=\sqrt{\dfrac{\mu_0\hbar\omega_c}{2V_c}}$ the magnetic field of vacuum, and $N_j$ the total number of spins in the sample. Plus, $\Omega=\dfrac{\gamma_e}{2}\sqrt{\dfrac{5\mu_0\rho_1 d_1D_p}{3c}}$ is the Rabi frequency of the applied drive, where $\rho_1$ and $d_1$ are the respective spin density and diameter of the first sample, while $D_p$ is incident power of the applied drive. To understand this transfer of spin-wave excitations at the level of transitions among energy levels, we note that in the absence of the extrinsic magnon drive, the Hamiltonian is excitation-preserving, with only an oscillatory energy transfer between the magnon and the cavity modes. Now let us label the eigenstates of the noninteracting system as $\ket{n_a,n_1,n_2}$, where $n_a$, $n_1$ and $n_2$ indicate populations of the cavity and the two magnon modes respectively. If a weak coherent drive (at the single-photon level) on $m_1$ excites the system into the state $\ket{0,1,0}$, energy would be exchanged back and forth among the energy levels  $\ket{0,1,0}$, $\ket{1,0,0}$, and $\ket{0,0,1}$, provided the dissipation is negligible. The transfer happens via the pathway $\ket{0,1,0}\to\ket{1,0,0}\to\ket{0,0,1}$, and this simple scheme can be easily extended to the case of coherent drives.

\begin{figure}
 \captionsetup{justification=raggedright,singlelinecheck=false}
 \centering
   \includegraphics[scale=0.5]{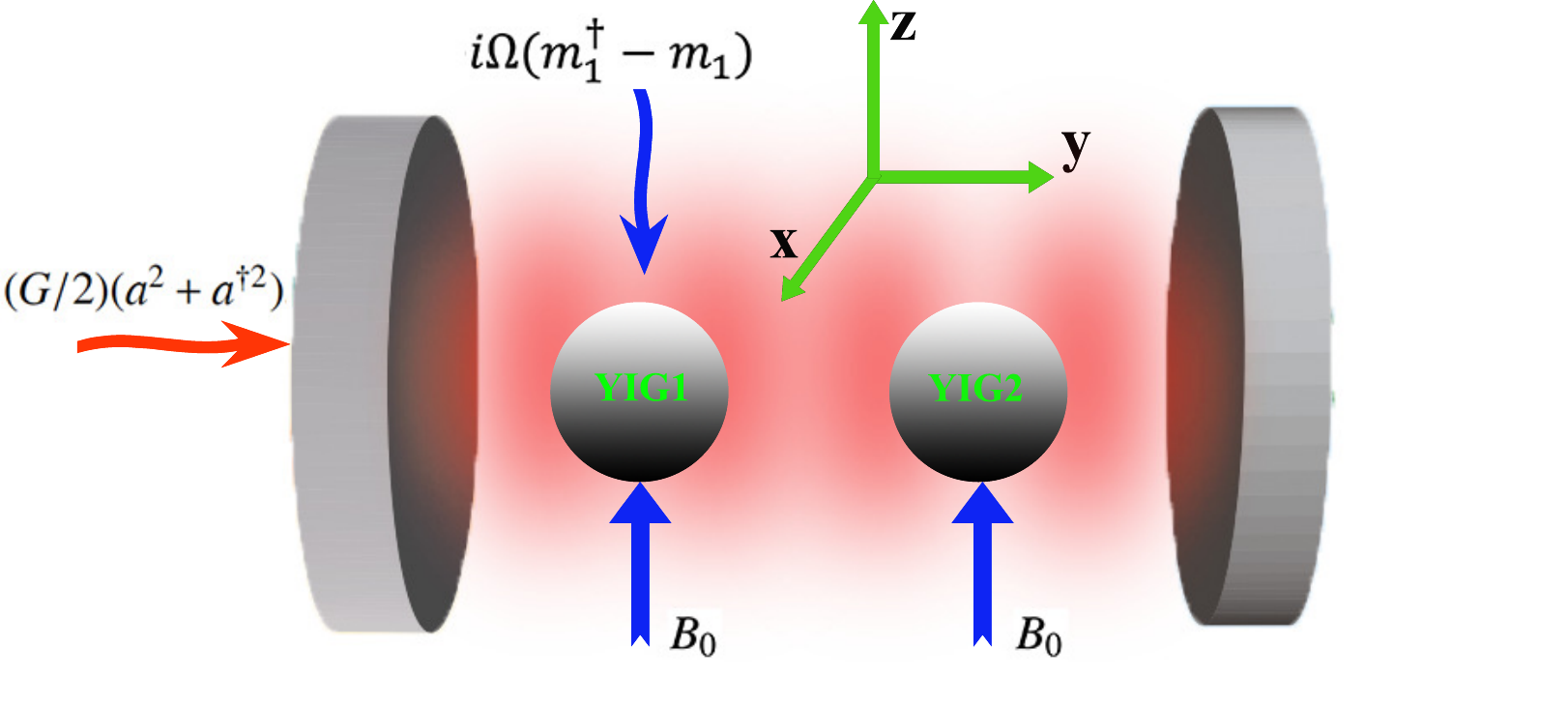}
\caption{Schematic of two ferrimagnetic samples of YIG, coherently coupled to a single-mode cavity, which is driven externally by a two-photon parametric drive. A uniform bias magnetic field $B_0$ applied to either of the YIG spheres generates the corresponding Kittel mode and the YIG1 is driven externally by a coherent drive of low photon occupancy.}
\label{sch}
\end{figure}

\textit{Amplified spin current:} We now demonstrate the impact of a parametric drive applied to the cavity on the associated transfer efficiency in the system considered above. Precisely, we would be leveraging the potential of parametrically enhanced spin-photon interactions to amplify the spin currents from magnetic samples loaded into a cavity resonator. The new schematic is portrayed in Fig. 1, where the cavity field is now parametrically driven. The preceding Hamiltonian has to be supplemented by an additional contribution of the form $\mathcal{H}_{\text{p}}/\hbar=(G/2)(a^2+a^{\dagger 2})$, so that the Hamiltonian for the parametric system \cite{PhysRevLett.120.093602, PhysRevLett.120.093601, zhu2020squeezed} would be given by
$\mathcal{H}=\mathcal{H}_0+\mathcal{H}_{\text{p}}$. To quell the time-dependence of the Hamiltonian, the frequency of the applied magnon drive $\omega_d$ has been set equal to $\omega_p/2$. Our objective is to investigate the steady-state spin-current response from the second YIG. Under the semiclassical approximation, the dynamical equations for the mode amplitudes at the level of mean fields can be obtained from the master equation of the system. These equations can be condensed in the form of a $6\cross 6$-matrix-differential equation
\begin{align}
\dot{X}=-iH_{\text{eff}}X+\Omega F_{\text{in}},
\end{align}
where $X=(a \hspace{1mm} m_1 \hspace{1mm} m_2 \hspace{1mm} a^{\dagger} \hspace{1mm} m_1^{\dagger} \hspace{1mm} m_2^{\dagger})^T$,  $H_{\text{eff}}=\begin{pmatrix}
H_{\text{eff}}^{(0)} & J\\
-J & -H_{\text{eff}}^{*(0)}
\end{pmatrix}$ is a $6\cross 6$ coupling matrix, and $F_{\text{in}}=(0 \hspace{1mm} 1 \hspace{1mm} 0 \hspace{1mm} 0 \hspace{1mm} 1 \hspace{1mm} 0)^T$. The expectation-value notations $\expval{.}$ have been dropped for brevity. The constituent block elements of $H_{\text{eff}}$ are given by $H_{\text{eff}}^{(0)}=\begin{pmatrix}
\Delta_c-i\kappa & g_1 & g_2\\
g_1 & \Delta_1-i\gamma_1 & 0\\
g_2 & 0 & \Delta_2-i\gamma_2\\
\end{pmatrix}$ and $J=\begin{pmatrix}
G & 0 & 0\\
0 & 0 & 0\\
0 & 0 & 0\\
\end{pmatrix}$, where $2\kappa,2\gamma_1,2\gamma_2$ are the respective relaxation rates of the cavity mode and the two magnon modes. Thus $H_{\text{eff}}^{(0)}$ denotes the coupling matrix in the absence of the parametric drive, i.e., with $G$ set equal to 0. Eq. (4) would permit a steady-state solution insofar as the eigenmodes of $H_{\text{eff}}$ have decaying character. Subject to this assumption, the steady-state spin-current response from the second magnetic sample could be expressed as $\mathscr{M}=\abs{m_2}^2$, wherein $m_2=\sum_{j=2,5}\bigg[-iH_{\text{eff}}^{-1}\bigg]_{2j}\Omega$.
 
\begin{figure}
 \captionsetup{justification=raggedright,singlelinecheck=false}
 \centering
   \includegraphics[scale=0.48]{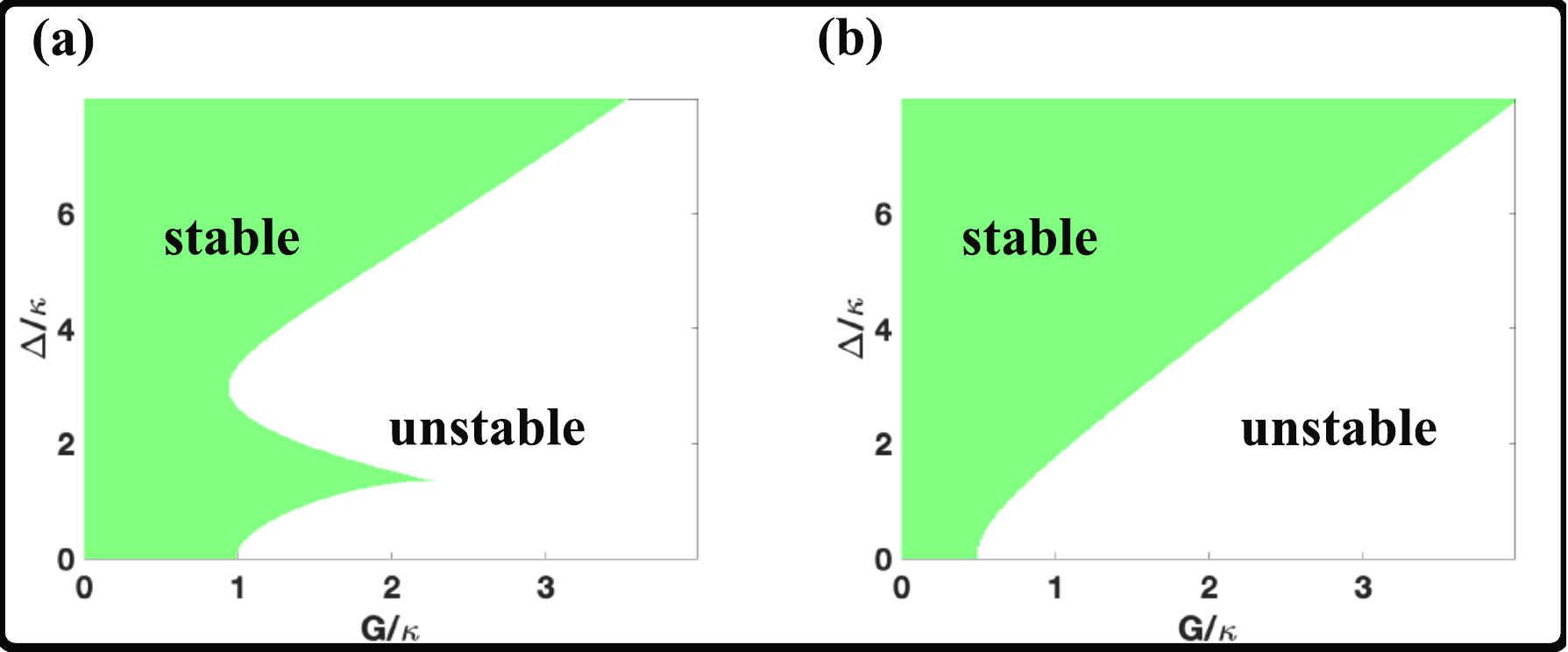}
\caption{Phase diagram highlighting the region of stable dynamics for (a) the three-mode cavity-magnon system considered in this letter, and (b) a parametrically driven single-mode cavity without any coupled accessories.}
\label{sch}
\end{figure}

To keep the analysis straightforward, we henceforth work with the assumption that $\Delta_1=\Delta_2=\Delta_c=\Delta$, $\kappa=\gamma_1=\gamma_2=\gamma$, and $g_1=g_2=g$. As was just stated, the stability of the steady state hinges on the imaginary parts of the eigenvalues, which we label as $\lambda_i$'s. This is formally equivalent to the Routh-Hurwitz criterion for the stability of nonlinear dynamics. The stable and the unstable regimes for this parametric model are numerically plotted in Fig. 2(a), for the ratio $g/\kappa=2$, with the demarcating partition between the two indentified by the equality $\Im(\lambda_i)=0$ for \textit{at least} one $\lambda_i$. Commensurately, there exists a critical parametric coupling strength $G_c(\Delta)$ below which the dynamics is stable. Contrast this with the corresponding stable regime obtained in the absence of magnons, i.e., $g=0$, which is juxtaposed for reference in Fig. 2(b). Not only does the magnon-photon coupling skew the stablity criterion, it does so in a highly nontrivial fashion. The effect is clearly not monotonic in $\Delta$. It can also be shown that the stability criterion does not depend on the two-photon detuning $\delta=\omega_d-\omega_p/2$ and the same phase diagram would apply to the case of non-zero detuning.

In order to capture the effect of the parametric drive, we introduce a dimensionless parameter $F$ that embodies the relevant enhancement in the spin-current response, $F=\frac{\mathscr{M}_{G\neq 0}}{\mathscr{M}_{G=0}}$. To start off, we note that this ratio is independent of the Rabi frequency $\Omega$, and therefore, insensitive to the incident photon pumping rate. In other words, the ratio $F$, for a specified value of $G$, is fundamentally predicated on the properties intrinsic to the system. Consequently, any departure of $F$ from unity is construed as a reinforcement due to the two-photon driving term.

\begin{figure}
 \captionsetup{justification=raggedright,singlelinecheck=false}
 \centering
   \includegraphics[scale=4]{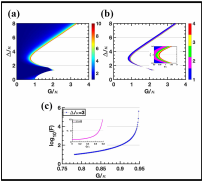}
\caption{Spin-current amplification factor plotted (a,b) over the two-dimensional $\Delta-G$ plane, and (c) against $G$ for $\Delta/\kappa=3$. The figures in (b,c), plotted in logarithmic scale, highlights the region where $F$ exceeds $10$, and testify to the tremendous level of enhancement near the phase boundary $G_c/\kappa=0.95$. The inset in (b) depicts a narrow segment of this region while the inset in (c) is a plot over the domain where $F$ is bounded above by $10$.}
\label{sch}
\end{figure}

We next present detailed results for the spin current based on Eq. (3). Note that we choose parameters such that the system is stable. Fig. 3(a,b) foregrounds the impact of variation in $G$ on $F$ in the form of a two-dimensional (2D) plot. The color-coded expanse within the graph, with $G<G_c$, pertains to the stable regime. Over a significant extent within the stable domain, the factor remains appreciably higher than unity. Particularly prominent is the narrow belt near the partition between the stable and unstable regimes, where the magnitude of $F$ starts surging expeditiously to larger orders of 10. In view of this, we have split the stable domain into two distinct regions. Fig. 3(a) corresponds to the region over which the amplification factor is bounded above by 10. For the range of parameters close to the stability partition, the logarithmic graph in Fig. 3(b) displays a significantly large spot where the magnitude of $F\gg10$. To provide useful estimates, a one-dimensional (1D) projection of the 2D plot onto the $\Delta/\kappa=3$ subspace is showcased in Figs. 3(c). Initially, as the narrow strip near the boundary is approached, we encounter about a tenfold enhancement in the spin-current response, i.e., $F\simeq\mathcal{O}(10)$. Further advances towards the boundary renders prodigious enhancements by several orders of magnitude ($F\simeq \mathcal{O}(10^2-10^4)$), evidenced by the inset in Fig. 3(c). What is intrigiuing is that these remarkable enhancement factors are all accessible within the stable regime. One may ask if it is possible to enhance the standard ($G=0$) photon mediated spin current by simply increasing the coupling strength $g$ or by ramping up the driving strength $\Omega$. However, this turns out not to be the case. The effect of increasing $g$ decreases the energy-transfer after hitting a peak intensity. Increasing the driving strength also does not help much since nonlinear effects set in and the transfer efficiency gets saturated, which is known from both theoretical \cite{PhysRevB.102.104415} and experimental \cite{PhysRevLett.120.057202} literature. For completeness, in Fig. 4, we also compare the enhancement factor for a non-zero magnitude of the two-photon detuning $\delta=\omega_d-\omega_p/2$ against the corresponding graph in the resonant setting. The inclusion of a two-photon detuning serves to regularize the somewhat divergent behavior of $F$ near the edge of the phase boundary. The detuning merely translates the coupling matrix as $H_{\text{eff}}\to {H}_{\text{eff}}-\delta$. On account of this rigid translation, the imaginary parts of the eigenvalues remain unaffected while the real parts get shifted by $-\delta$. This explains why the phase boundary does not change but the almost divergent nature of the response near the boundary is further stabilized. We add that the techniques for producing parametric drives in microwave domain are known from the early work of Yurke \cite{PhysRevLett.65.1419} and many others \cite{castellanos2008amplification, PhysRevLett.105.233907}. Josephson junction amplifiers operating at low temperatures have evolved as the keystone for these experiments. This is in contrast to the ones in the optical domain which work at room temperature and utilize second order nonlinearities in crystals (see, for example, \cite{PhysRevLett.101.233602}). As an alternative strategy sustainable at room temperature, we also float the possibility of modulating the parametric pump at twice the cavity frequency, thereby generating the desired parametric interaction. 

\begin{figure}
 \captionsetup{justification=raggedright,singlelinecheck=false}
 \centering
   \includegraphics[scale=0.35]{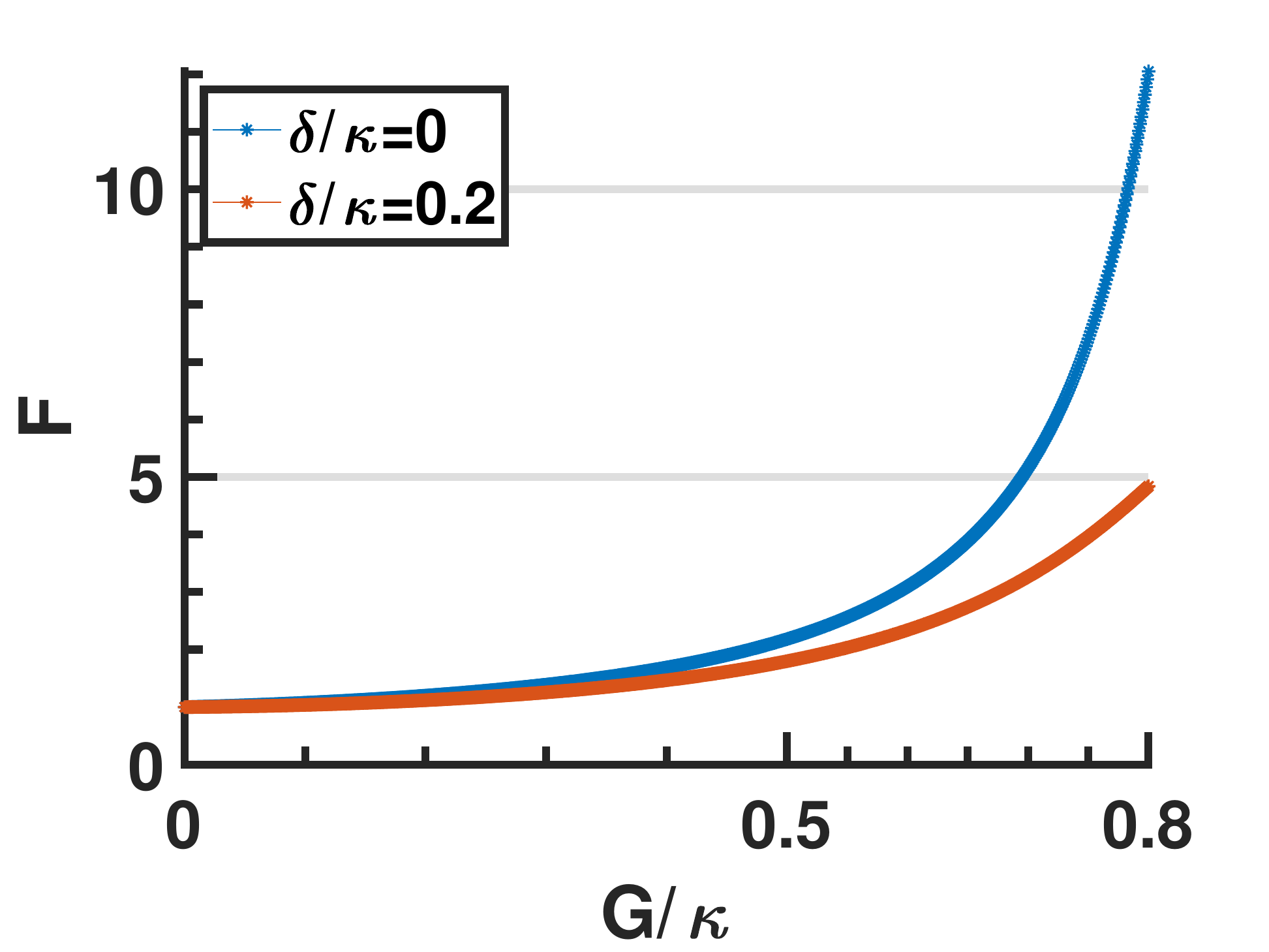}
\caption{Comparative graph illustrating the respective enhancement factors under the two-photon resonance condition $\delta/\kappa=0$, and in the detuned scenario, with $\delta/\kappa=0.2$ chosen as an example. Here, $\delta=\omega_d-\omega_p/2$ signifies the two-photon detuning.}
\label{sch}
\end{figure}

While we have presented the numerical results of the spin-current enhancement using the semiclassical approach, it is worthwhile to examine the contributions from quantum fluctuations to these semiclassical results. To account for quantum corrections, we need to work with the quantum Langevin equations \cite{PhysRevLett.121.203601} and evaluate the covariance matrix of photon and magnon variables. The analysis demonstrates that the quantum contribution remains many orders of magnitude smaller than the semiclassical contribution and can be ignored. The physical reason behind this is the the Langevin forces are delta-correlated, i.e. $\expval{F(t)F^{\dagger}(t')}=2\gamma\delta(t-t')$, so that the fluctuations are effectively driven by a flux of strength $\gamma$. The semiclassical response, on the contrary, ensues from a classical driving field characterized by the Rabi frequency $\Omega$. Since for a driving power of $1$ $\mu$W, $(\Omega/\gamma)\sim 10^5$, the quantum contribution remains strongly subordinated.

\begin{figure}
 \captionsetup{justification=raggedright,singlelinecheck=false}
 \centering
   \includegraphics[scale=0.48]{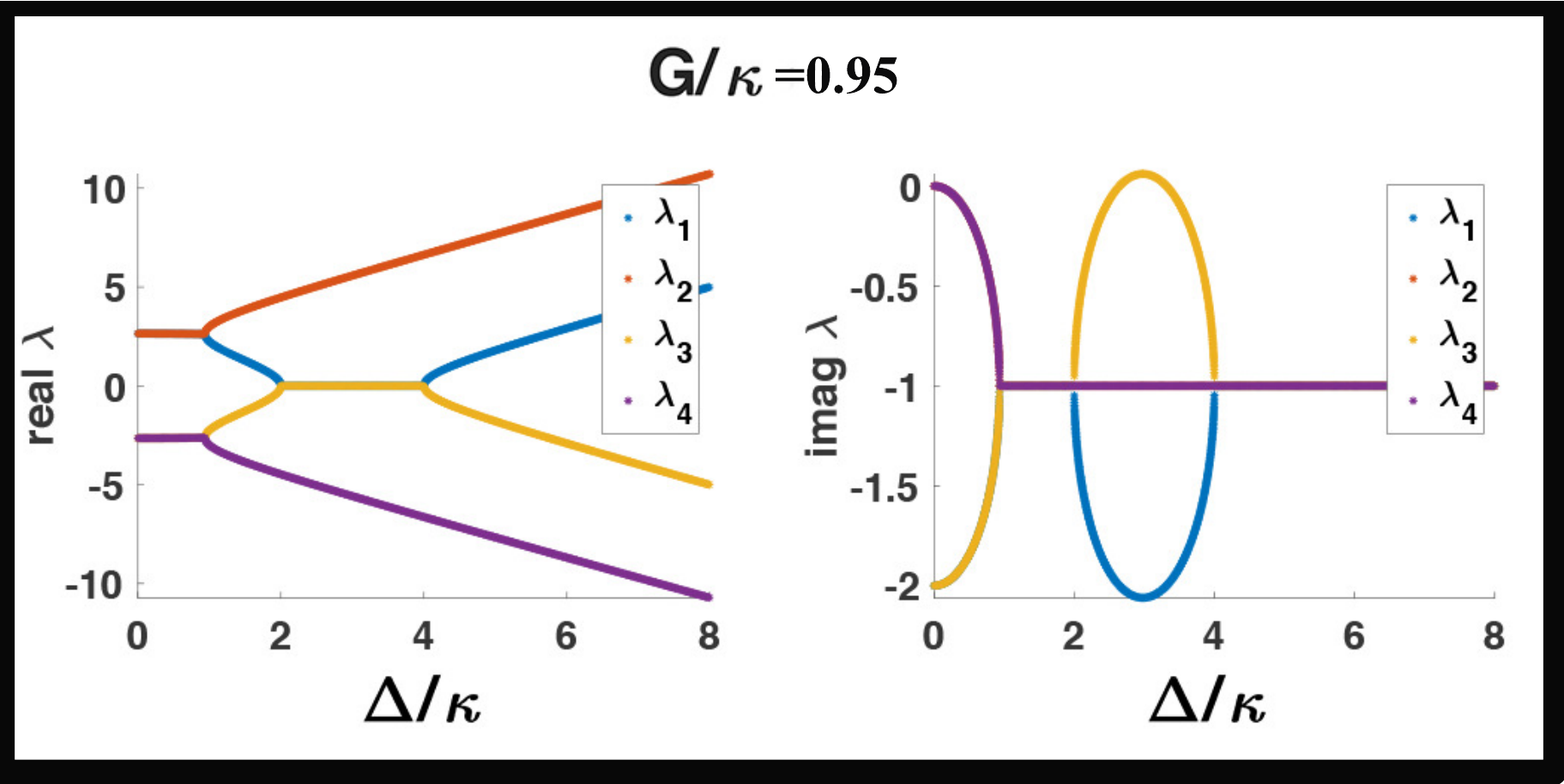}
\caption{(a) Real and (b) Imaginary parts of the eigenvalues of the reduced $4\cross 4$ system involving the modes $a$ and $M=(m_1+m_2)/\sqrt{2}$ (as described by Eq. (10)).}
\label{sch}
\end{figure}

\textit{Parametric-drive-induced long-lived mode:} The anomalously large enhancement observed along the inner fringes of the stability boundary, and depicted in Fig. 3(c), is an immediate consequence of parametrically induced magnon-photon coherences. In particular, strong coherence leads to the emergence of a long-lived mode which sharply reinforces the response function. We can streamline the characterization of eigenvalues by expressing the starting Hamiltonian in terms of the collective magnon operators $M=(m_1+m_2)/\sqrt{2}$ and $m=(m_1-m_2)/\sqrt{2}$. This leads to
\begin{align}
\mathcal{H}/\hbar=\Delta a^{\dagger}a+\Delta M^{\dagger}M+\Delta m^{\dagger}m+g[a^{\dagger}M+aM^{\dagger}]\notag\\+G(a^2+a^{\dagger 2})+i\Omega[(M^{\dagger}+m^{\dagger})-\text{h.c})],
\end{align}
Clearly, the mode $m$, which is decoupled from the cavity field, can be dispensed with, and the coupled dynamics would involve only $a$, $M$, and their adjoints. The relevant dynamical equation can now be cast as
\begin{align}
\dot{\mathscr{X}}=-i\mathscr{H}_{\text{eff}}\mathscr{X}+\Omega \mathscr{F}_{\text{in}},
\end{align}
where we have $\mathscr{X}=(a \hspace{1mm} M \hspace{1mm} a^{\dagger} \hspace{1mm} M^{\dagger} )^T$ and $\mathscr{H}_{\text{eff}}=\begin{pmatrix}
\mathscr{H}_{\text{eff}}^{(0)} & \mathscr{J}\\
-\mathscr{J} & -\mathscr{H}_{\text{eff}}^{*(0)}
\end{pmatrix}$, with $\mathscr{H}_{\text{eff}}^{(0)}=\begin{pmatrix}
\Delta-i\gamma & g\\
g & \Delta-i\gamma
\end{pmatrix}$, and $\mathscr{J}=\begin{pmatrix}
2G & 0\\
0 & 0
\end{pmatrix}$. We now plot the four eigenvalues (real and imaginary parts) of the reduced Hamiltonian $\mathscr{H}_{\text{eff}}$ at $G/\kappa=0.95$ in Figs. 5(a,b). It is observed that there exists a particular eigenvalue $\lambda_3$ with vanishing real part, and  whose imaginary part becomes tantalizingly close to zero at $\Delta/\kappa\approx 3$. Viewed alternatively, the mode identified by $\lambda_3$ is symbolic of a long-lived or quasi-dark mode in the system. Since $\mathscr{H}_{\text{eff}}^{-1}\sim (\det[\mathscr{H}_{\text{eff}}])^{-1}\sim(\prod_{j}\lambda_j^{-1})$, the effect of a vanishingly diminutive eigenvalue is reflected in the dramatic amplification of the steady-state response. The stark enhancement in the spin current stems from the energy of a two-photon parametric drive and stands testimony to the emergence of higher-order energy transition pathways ensuing from the parametric term (see Fig. 6). Without the parametric drive, the transfer efficiency was dictated by the applied magnon drive. However, with the provision of a parametric drive, an elementary two-photon excitation can raise the state $\ket{1,0,0}$ to $\ket{3,0,0}$. As the intracavity field interacts with the magnons, this energy can be redistributed amongst all the participating entities, leading to a myriad of allowed final states, such as $\ket{2,1,0}$, $\ket{1,1,1}$, and $\ket{1,0,2}$, to cite a few. Therefore, the possibility of effecting higher-order magnon occupancies opens up, even in the limit of very weak pumping rates. In general, for adequately strong values of $G$, transitions from $\ket{1,0,0}$ to higher-order excited states of the form $\ket{2n+1,0,0}$ can be achieved with greater probabilities, which, in turn, would engender higher-order excitations of the magnon modes. While we illustrate the energy transitions assuming an external drive at the single-photon level, these arguments are easily generalized to the case of larger coherent drives. 

To sum things up, we have brought to light the emergence of parametrically induced strong coherence effects in a cavity-QED system which can lead to magnificent enhancements in the cavity-mediated transfer of spin excitations. In light of the recent upswing of interest in cavity magnonics, we have highlighted the application of this principle to the parametric amplification of spin currents in magnetic samples. To underline the origin of this enhancement, we have set forth a compact explanation in terms of higher-order energy transitions from the system's ground state stimulated by the cascaded two-photon absorption from the parametric pump. Increasing the strength of the parametric drive only serves to embolden this effect, with progressively higher and higher spin currents being observed. The efficiency of this process peaks at the two-photon resonance condition $\omega_d=\omega_p/2$, supporting enhancement by orders of magnitude even within the stable phase of the system dynamics. The spectacular amplification of spin-current transfer at the resonance condition stems from a parametrically produced long-lived mode in the system. The developed formalism and the accompanying results would apply to various kinds of systems. Quite generally, the energy-transfer protocol could be applied to systems of atoms, superconducting qubits and quantum dots, trapped ions, and spin ensembles, in general. Likewise, nonlinear effects such as bistability would also be reinforced on account of the parametrically induced coherences. 

\begin{figure}
 \captionsetup{justification=raggedright,singlelinecheck=false}
 \centering
   \includegraphics[scale=0.19]{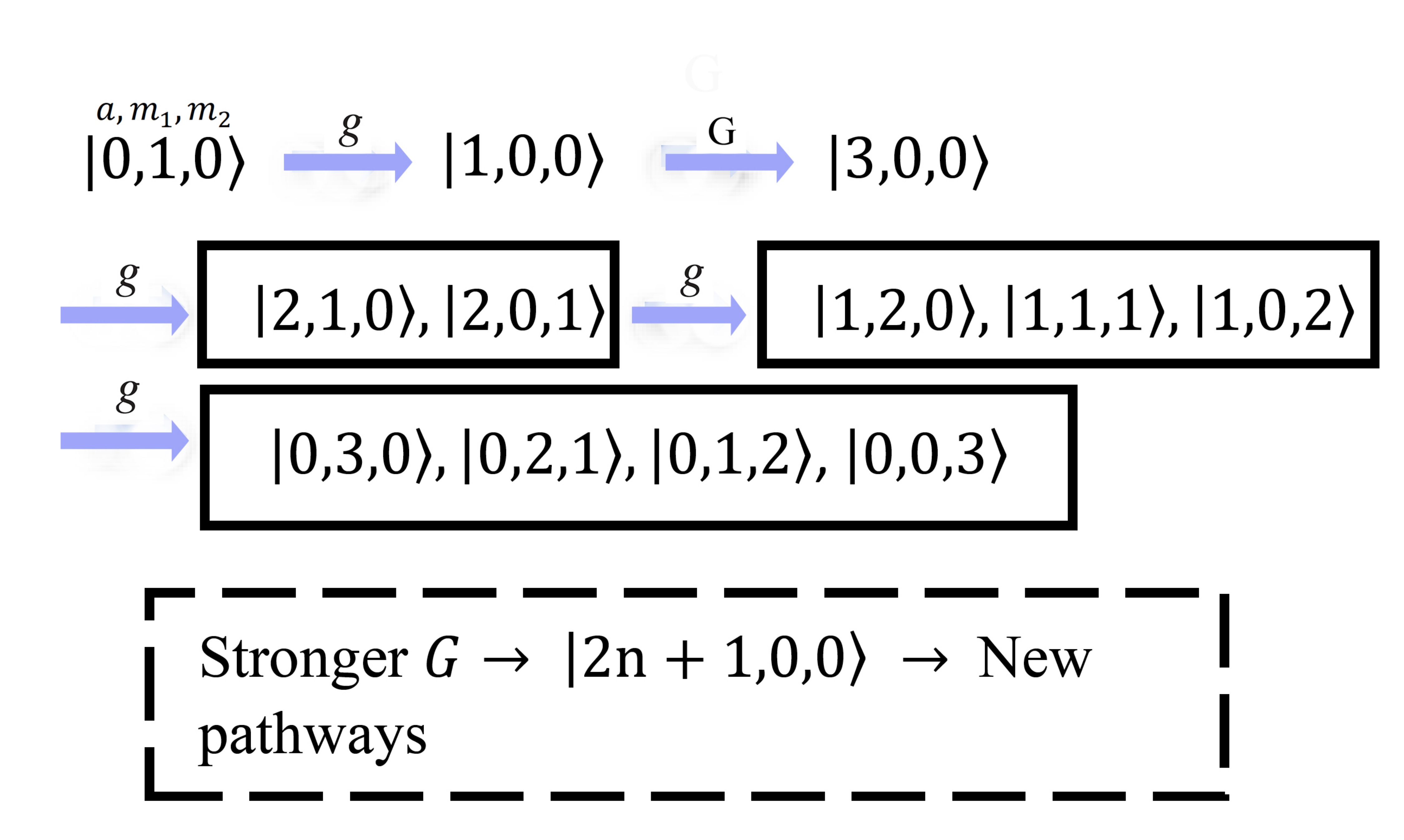}
\caption{Higher-order energy transitions afforded by the parametric drive, assuming that YIG1 is driven by a coherent field at one-photon level.}
\label{sch}
\end{figure}

\section{Acknowledgments}
The authors gratefully acknowledge the support of The Air Force Office of Scientific Research [AFOSR award no. FA9550-20-1-0366], The Robert A. Welch Foundation [grant no. A-1243] and the Herman F. Heep and Minnie Belle Heep Texas A$\&$M University endowed fund.

J.M.P.N. and D.M. contributed equally to this work. 
\bibliography{references}
\end{document}